\documentclass[aps,pra,twocolumn,showpacs]{revtex4}
\usepackage{mathrsfs}
\usepackage{graphicx}
\usepackage{amsmath}
\usepackage{latexsym}
\usepackage{amsfonts}
\usepackage{amssymb}
\usepackage{array}
\usepackage{epsfig}
\newcommand{\bra}[1]{\left\langle{#1}\right\vert}
\newcommand{\ket}[1]{\left\vert{#1}\right\rangle}

\newcommand{\ketbra}[2]{|#1\rangle \langle#2|}
\newcommand{\expect}[3]{\langle#1|#2|#3\rangle}
\newcommand{\ovl}[2]{\langle #1 | #2 \rangle}

\newcommand{\ncd}{\newcommand}
\newcommand{\be}{\begin{equation}}
\newcommand{\ee}{\end{equation}}
\newcommand{\ba}{\begin{array}}
\newcommand{\ea}{\end{array}}
\newcommand{\bqa}{\begin{eqnarray}}
\newcommand{\eqa}{\end{eqnarray}}
\newcommand{\ie}{{\it i.e. }}
\newcommand{\tr}{\mbox{Tr}}

\ncd{\SABC}{S^{ABC}}
\ncd{\Sab}{S_{ab}}
\ncd{\Sbc}{S_{bc}}
\ncd{\Sba}{S_{ba}}
\ncd{\csk}{{|\phi_{\{\kappa\} }
\rangle}_{\cal{C}}}
\ncd{\nbgh}{\text{nghb}}
\ncd{\QC}{$\mbox{QC}_{\cal{C}}$}

\begin{document}
\title{Decoherence-based exploration of $d$-dimensional one-way quantum computation}
\author{M. S. Tame$^1$, M. Paternostro$^{1,\footnote{Present address: Institute for Quantum Optics and Quantum Information (IQOQI), Austrian Academy of Sciences, Boltzmanngasse 3, A-1090 Vienna, Austria}}$, C. Hadley$^2$, S. Bose$^2$, and M. S. Kim$^1$}
\affiliation{$^1$School of Mathematics and Physics, The Queen's University, Belfast BT7 1NN, UK}
\affiliation{$^2$Department of Physics and Astronomy, University College London, Gower Street, London WC1E 6BT, UK}
\date{\today}
\begin{abstract}
We study the effects of amplitude and phase damping decoherence in $d$-dimensional one-way quantum computation (QC). Our investigation shows how information transfer and entangling gate simulations are affected for $d\ge{2}$. To understand motivations for extending the one-way model to higher dimensions, we describe how $d$-dimensional {\it qudit} cluster states deteriorate under environmental noise. In order to protect quantum information from the environment we consider the encoding of logical qubits into physical qudits and compare entangled pairs of linear qubit-cluster states with single qudit clusters of equal length and total dimension. Our study shows a significant reduction in the performance of one-way QC for $d>2$ in the presence of Markovian type decoherence models.
\end{abstract}
\maketitle
\section{Introduction}
The one-way model for quantum computation (QC)~\cite{RBH} is an appealing alternative to the standard quantum circuit approach for physical systems where multipartite entangled resources, known as graph states~\cite{HEB}, can be generated with a minimal amount of dynamical processes. A certain class of these states, known as {\it cluster states}, have proven to be useful as universal resources upon which adaptive measurement based QC can be carried out. Recently, considerable attention has been focused on cluster state based QC, in both theoretical~\cite{Hein} and experimental contexts~\cite{Wal}. 

The standard one-way model relies on the use of entangled two-dimensional systems (qubits) and adaptive single-qubit measurements to propagate information and simulate quantum gates. Recently, this model was extended to $d$-dimensional {\it qudit} systems~\cite{Zhou1}. Many physical setups exist that could be tailored to embody systems with the correct entanglement structure for qudit cluster states. These include ion-traps~\cite{Klimov}, cavity quantum electrodynamical (cavity-QED) settings~\cite{Zheng} and linear optical setups~\cite{Vaz}. Compared to qubits, $d$-dimensional systems ($d\ge3$) provide improvements in channel capacities for quantum communication~\cite{Fuj}, better levels of security in quantum bit-commitment and coin-flipping protocols~\cite{Spek} and violations of local realism are much stronger for two maximally entangled qudits~\cite{Kas}. Three-dimensional systems (qutrits) are also known to optimize the Hilbert space dimensionality for QC power \cite{Green}. However, so far it is not clear if the use of $d$-dimensional information carriers provides any substantial advantage in one-way QC. It is therefore interesting to investigate the use of $d$-dimensional systems in this context to see if advantages can be given by accessing a larger Hilbert space. The accuracy of quantum information processing (QIP) protocols using qubit cluster states is known to be affected significantly by sources of environmental decoherence and imperfections~\cite{Tame1} and removing all but only the most essential qubits in the cluster state is key to effective simulations~\cite{Nielsen1}. The central aim of this work is the study of the behavior of QIP carried out when an environment affects single-mode $d$-dimensional systems that comprise the qudit cluster states. We find that disadvantages appear, in terms of robustness of QIP protocols, when moving to higher dimensions; the accuracy of simulations decreases as the dimension increases. In addition, entangled pairs of qubit linear clusters appear to be more resilient to environmental effects in comparison to qudit clusters. Our study questions the worthiness of efforts made toward the extension of one-way QC to higher-dimensional systems, where global properties of the entangled resources are crucial for the performance of a given QIP protocol.

In Section II we provide an overview of $d$-dimensional one-way QC. In Section III, we introduce the decoherence models used in our analysis and determine their overall effect on qudit cluster states using the state fidelity. Entanglement decay is also studied using techniques for deducing concurrence in bipartite mixed states of arbitrary dimension~\cite{Mintert2}. However, it is not straighforward to compare properties such as the strength of entanglement or fidelity for states from different dimensions. Thus, in Section IV we take an operational point-of-view and focus attention on the performance of information transfer, gate simulations and encoding techniques. Section V summarizes our results. 
\section{Background}
A qudit cluster state $|\phi \rangle_{{\cal{C}}}$ is a pure multipartite entangled state of qudits positioned at specific sites of a lattice structure known as the cluster ${\cal{C}}$. It is defined as the eigenstate of the set of operators \cite{RBH,Zhou1} $ K^{(a)} =  X^{\dag}_{a}\bigotimes_{b}Z_{b}$, where $X$ and $Z$ are generalized Pauli operators \cite{Knill} given by $Z = \sum_{k=0}^{d-1} \omega^k \ketbra{k}{k}$ and $X = \sum_{k=0}^{d-1} \ketbra{k-1}{k}$. Modulo-$d$ arithmetic is used for $k$, $\omega=e^{2 \pi i/d}$ is the $d$-th root of unity and $\{ \ket{k} \}_{k=0}^{d-1}$ is a basis of eigenstates of $Z$ with eigenvalues $\omega^{k}$. Each $K^{(a)}$ acts on the qudit occupying site $a\in{\cal{C}}$ and all others occupying a neighboring lattice site $b$. The cluster state $|\phi \rangle_{{\cal{C}}}$ can be generated by first preparing a product state $|+\rangle_{\cal{C}}= \bigotimes_{a
\in {\cal{C}}} |+\rangle_{a}$ of the qudits at all sites $a$, where the Fourier transform basis $\ket{+_j} = 1/\sqrt{d}\sum_{k=0}^{d-1} \omega^{jk} \ket{k}$ is used with $|+\rangle:=|+_0\rangle$. The set $\{ \ket{+_j} \}_{j=0}^{d-1}$ contains the eigenstates of the operator $X$ with eigenvalues $\omega^{j}$ respectively. A unitary transformation $S^{({\cal{C}})}=\prod_{\langle a,b\rangle}\Sab$ is then applied to the initial state $|+\rangle_{\cal{C}}$, where $\langle a,b\rangle:=\{a,b\in{\cal{C}}|b-a\in\gamma_D\}$ and $\gamma_1=\{1\}$, $\gamma_2=\{(1,0)^T,(0,1)^T\}$, $\gamma_3=\{(1,0,0)^T,(0,1,0)^T, (0,0,1)^T \}$ for the respective spatial dimension
$D$ of the cluster being used. Each $\Sab$ can be described by the entangling operator~\cite{Zhou1}
\begin{equation}
\label{Sabdef}
\Sab= \sum_{k=0}^{d-1}|k \rangle_a \langle k| \otimes Z_b^k=\sum_{k,l=0}^{d-1} \omega^{kl} |k,l \rangle_{a,b} \langle k,l| .
\end{equation}
The state generated by the action of $S^{({\cal{C}})}$
on $|+\rangle_{\cal{C}}$ is found to be
$S^{({\cal{C}})} |+\rangle_{\cal{C}}\equiv\prod_{\langle a,b \rangle}
\Sab \bigotimes_{a \in {\cal C}}|+\rangle_{a}=|\phi \rangle_{{\cal{C}}}$,
where the cluster state $|\phi \rangle_{{\cal{C}}}$ satisfies the eigenvalue equations
$K^{(a)}|\phi \rangle_{{\cal{C}}}=|
\phi \rangle_{{\cal{C}}},~\forall a \in {\cal C}$.
In order to carry out quantum simulations, a cluster of qudits in a particular physical configuration ${\cal{C}}(g)$ is used. To understand how to design correct configurations for carrying out specific protocols, it is convenient to start from the concept of qudit Basic Building Blocks (BBB's) and their equivalent network circuits. We then use simple concatenation rules to build up more complicated protocols, in a similar way to the qubit case~\cite{Tame1}. For a short summary, see Appendix A.
\section{Decoherence models and general properties}
\begin{figure}[t]
\centerline{
\psfig{figure=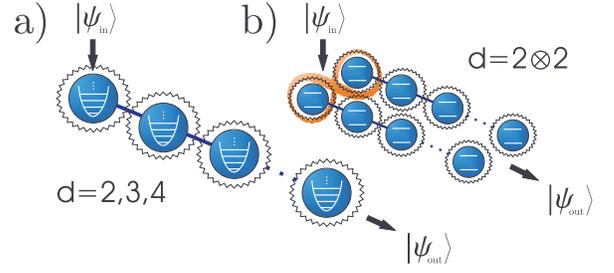,width=7.7cm}}
\caption{The linear qudit-clusters studied in this work. Each physical qudit is affected by its local environment (jagged surroundings) by AD and PD decoherence. {\bf (a)}: Linear qudit clusters. {\bf (b)}: Entangled pair of linear qubit-clusters.}
\label{figu1a}
\end{figure}

In the analysis presented here, we consider each physical qudit in the cluster state interacting independently with a local environment as shown in Fig. \ref{figu1a} {\bf (a)}. The evolution of the state of a single qudit can be given in terms of the master equation $\frac{{\rm d} \varrho}{{\rm d}t}={\cal L}_D [\varrho]$, where $\varrho$ is the density matrix of the qudit and ${\cal L}_D$ represents the Lindblad operator describing the particular decoherence model. In this work, we will treat each qudit as a bosonic mode with a truncated basis of length $d$. Taking the local environment as a thermal bath, one may write the Lindblad operator acting on the qudit as
\begin{equation}
\label{thermal}
\begin{split}
{\cal L}_{A}[\varrho]=g^{1}_A(a\varrho{a}^\dag-a^\dag{a}\varrho)+g^{0}_A(a^\dag\varrho{a}-a{a^\dag}\varrho)+h.c
\end{split}
\end{equation}
where $g^{k}_A=({\Gamma_A}/{2})(\bar{n}+k)$, $\Gamma_A$ is the strength of the qudit-environment coupling, $a^{\dag}~(a)$ denotes the creation (annihilation) operator for the qudit and $\bar{n}$ parameterizes a non-zero temperature environment. This model is usually referred to as amplitude damping (AD) and characterizes the energy dissipation of a system to its environment.
If the local environment acts to destroy the phase-coherence in the qudit state via random scattering processes, the Lindblad operator can be written as 
\begin{eqnarray}
\label{dep}
{\cal L}_{P} [\varrho] = \frac{\Gamma_P}{2}  \left(2\, a^{\dag} a \varrho a^{\dag} a -
\{(a^{\dagger} a)^2, \varrho\} \right).
\end{eqnarray}
Here the rate $\Gamma_P$ represents the strength of the scattering process. This type of decoherence is usually referred to as phase damping (PD).
The master equations corresponding to the Lindblad operators in Eqs.~(\ref{thermal}) and (\ref{dep}) can be solved by expressing them in the single-qudit computational basis. 
However, in order to solve the dynamics of environment-affected many-qudit cluster states, it is convenient to rely on the Kraus operator formalism~\cite{Kraus}. In this context, we can write the evolution of a single qudit density matrix $\varrho$ as $\varrho(t)=\sum_{\mu}K_{\mu}^{i}(t)\varrho(0)K_{\mu}^{i}\mbox{}^{\dag}(t)$, where $\{K_{\mu}^{i} (t)\}$ is the set of Kraus operators for qudit $i$ satisfying the relation $\sum_{\mu}K_{\mu}^{i}\mbox{}^{\dag}(t) K_{\mu}^i(t)=\openone,(K=A,P)$. The AD Kraus operator is given by
\begin{equation}
A^i_{\mu}(t)=\sum_{n=\mu}^{\infty}[C(n,\mu) \gamma(t)^{n-\mu}(1-\gamma(t))^{\mu}]^{\frac{1}{2}}|n-\mu \rangle_i \langle n |,\label{A}
\end{equation}
where $(1-\gamma(t))^{\frac{\mu}{2}}$ is the probability that the qudit system loses $\mu$ particles up to time $t$ \cite{ChuangMilburn}. We set $\gamma(t)=e^{-\Gamma_A t}$ for the solution $\varrho (t)$ to be consistent with Eq. (\ref{thermal}) in the limit $\bar{n}=0$. The PD Kraus operator is given by \cite{Liu, Amosov}
\begin{equation}
P^i_{\mu}(t)=\sum_{n=0}^{\infty}e^{-\frac{1}{2}n^2\tau}
[\left(n^2\tau \right)^{\mu}/\mu!]^{\frac{1}{2}} |n\rangle_i\langle
n|\label{P}
\end{equation}
where $\tau=\Gamma_P t$ is chosen as a rescaled interaction time and $(1-e^{-n^2\tau})^{1/2}$ can be interpreted as the
probability that $n$ particles from the qudit system are scattered by
the environment. 
\begin{figure}[t]
\centerline{
\psfig{figure=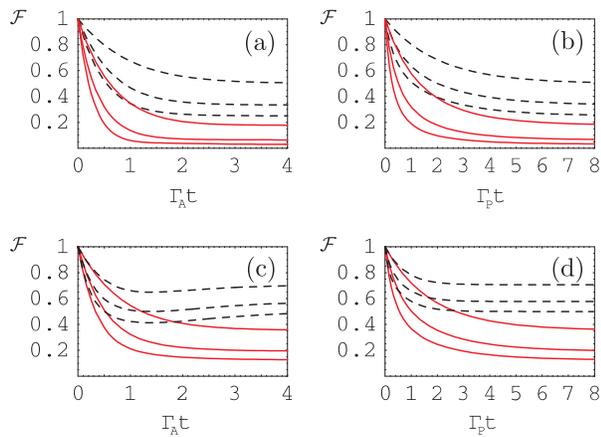,width=7.9cm}}
\caption{Fidelities of decoherence-affected linear qudit-clusters. In {\bf (a)} and {\bf (b)} the dashed (solid) lines correspond to $n=2$ ($n=5$) with dimension $d=2 \to 4$ from top to bottom in each line-style. In {\bf (c)} and {\bf (d)} we compare $n=3$ qudit cluster states (solid lines) with 3-qudit GHZ states (dashed lines) for $d=2 \to 4$ from top to bottom in each line-style. We consider AD ({\bf (a)} \& {\bf (c)}) and PD channels ({\bf (b)} \& {\bf (d)}).}
\label{figu1}
\end{figure}
For an $n$-qudit cluster state initially described by the density matrix $\varrho(0)$ and each qudit interacting with its own local environment, we then have an evolution described by $\varrho(t)=\sum_{\mu_i}\tilde{K}_{\{\mu_i\}}^{}(t)\varrho(t)\tilde{K}^\dag_{\{\mu_i\}}(t)$,
where $\tilde{K}_{\{\mu_i\}}(t)=\otimes^n_{i=1}K^i_{\mu_i}(t)$ and each $K_{\mu_i}^i(t)$ acts separately on qudit $i$ in the cluster. By truncating the basis at dimension $d$, we remove the infinity limit in the definition of $\{ A^i_{\mu}(t) \}$ as number states outside the $d$-dimensional Hilbert space do not play a role in the decoherence mechanism. The index $\mu$ is also restricted and results in a finite number of AD Kraus operators. For $\{ P^i_{\mu}(t) \}$, the index $\mu$ is not restricted, resulting in an infinite number of operators. This is because the system can be scattered by any number of particles in the environment. However, the index $n$ is restricted by the truncation of the basis at dimension $d$ and therefore one can redefine $\{ P^i_{\mu}(t) \}$ into a finite sum of PD Kraus operators~\cite{Liu}.
In order to give a general idea of how qudit cluster states are affected by both these decoherence models, we use the state fidelity given by ${\cal F}(\ket{\psi},\varrho)=\sqrt{\bra{\psi}\varrho \ket{\psi}}$ for a pure cluster state $\ket{\psi}$ and environment exposed mixed cluster state $\varrho$. In Fig.~\ref{figu1} the fidelities are shown for linear qudit cluster states of length $n=2,3$ and $5$ for dimensions $d=2,~3$ and $4$. In Fig.~\ref{figu1} {\bf (a)} and {\bf (b)}, we show AD and PD behavior for lengths $n=2$ and $5$. Here one can see that the higher the dimension of the qudits, the stronger the decay becomes with increased exposure time for both decoherence models.  
In Fig.~\ref{figu1} {\bf (c)} and {\bf (d)}, we compare $n=3$ qudit cluster states with their local unitary (LU) equivalent $n=3$ qudit GHZ states. Due to the basis used in order to express the cluster states (spread out across the eigenstate basis of the $Z$ operator), their fidelities decay more strongly than the GHZ states.

\begin{figure}[t]
\centerline{
\psfig{figure=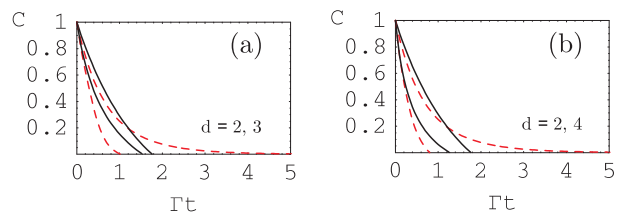,width=8cm}}
\caption{Bipartite entanglement decay in two-qudit cluster states when decoherence affects the individual qudits. The concurrence of qubit cluster states affected by AD and PD (top solid and dashed lines respectively) is compared with the normalized quasi-concurrence for $d=3$ {\bf (a)} and $d=4$ qudit cluster states {\bf (b)}. AD and PD correspond to the lower solid and dashed lines in each panel respectively.} 
\label{figu2}
\end{figure}

In addition to the state fidelity behavior we consider how decay in entanglement is affected as the dimension of the cluster increases. For the moment we limit the analysis to $n=2$ qudit cluster states, which are locally equivalent to the maximally entangled bipartite states $\ket{\Psi_d}=(1/\sqrt{d})\sum_{i=0}^{d-1}\ket{ii}$ for a given dimension $d$. Later, we consider bipartite entanglement decay in $n=3$ qudit cluster states where one of the qudits has been measured. For $d$-dimensional bipartite pure states $\varrho=\ket{\Psi}\bra{\Psi}$, the concurrence $c(\varrho)$~\cite{Wooters, Mintert4} provides a measure of entanglement. 
In general, the calculation of $c(\varrho)$ is a formidable task. However, it is possible to obtain approximations providing tight upper and lower bounds to $c(\varrho)$~\cite{Mintert4}. Here, we use the {\it quasi-pure} concurrence $c_{qp}(\varrho)={\max}(\lambda_1-\sum_{i>1}\lambda_i,0)$~\cite{Mintert2} with $\lambda_i$'s the eigenvalues of the matrix $\sqrt{\tau \tau^{\dag}}$ (decreasingly ordered) and $\tau_{jk}={\cal A}_{jk}^{11}/\sqrt{{\cal A}_{11}^{11}}$, ${\cal A}_{jk}^{11}=\mu_1\sqrt{\mu_j\mu_k}[
\tr(\ket{\Phi_j}\ovl{\Phi_1}{\Phi_k}\bra{\Phi_1})-\sum^2_{l\neq{m}=1}\tr_l(\tr_m(\ket{\Phi_j}\bra{\Phi_1})\tr_m(\ket{\Phi_k}\bra{\Phi_1})-\tr(\ket{\Phi_j}\bra{\Phi_1})\tr(\ket{\Phi_k}\bra{\Phi_1})]$. The set $\{\mu_i, \ket{\Phi_i}\}$ corresponds to the eigensystem of $\varrho$. This entanglement measure is ideal for describing the entanglement decay in a system where the environment acts to destroy its purity slowly. Under these conditions $c_{qp}(\varrho)$ represents a value very close to the actual concurrence $c(\varrho)$, with the approximation valid for $\mu_1 \gg \mu_{i>1}$ (the $\mu_i$ being non-increasingly ordered). When the approximation is no longer valid $c_{qp}(\varrho)$ nevertheless represents a lower bound to $c(\rho)$. In Fig.~\ref{figu2} we show the effect of AD and PD on entanglement decay in the form of the concurrence $c(\varrho)$ for $d=2$ and normalized quasi-concurrence $\tilde{c}_{qp}(\varrho)=c_{qp}(\varrho)/c(\Psi_d)$ for $d=3$ and $4$, where $c(\Psi_d)=\sqrt{2(1-1/d)}$. In Figs.~\ref{figu2} {\bf (a)} and {\bf (b)} we compare qubits with $d=3$ and $d=4$ systems respectively. In both decoherence models considered, we find that the quasi-concurrence decay is faster for larger dimension. However, we cannot infer that the total amount of entanglement decreases faster at higher dimensions. Only the fraction of the maximal value $c(\Psi_d)$ decays faster. We have checked the validity of the quasi-pure approximation
by inspecting the largest $\mu_1$ and second largest $\mu_2$ eigenvalues of the eigensystem decomposition of the decayed state $\varrho$.
\section{Manipulating Information}
\subsection{Information transfer} 
In this section we consider linear qudit cluster states of length $n=2 \to 5$ subject to the decoherence models of AD and PD introduced in the last section. Individual qudits in the clusters are exposed to a local environment for a rescaled interaction time $\Gamma_i t,\,i={\rm A,P}$. A logical state is encoded on the first physical qudit and measurements are performed in order to propagate the state across the cluster: see Fig.~\ref{figu1a} {\bf (a)}. This simple model gives an idea of how information flow is affected in general and the range of lengths of clusters considered allows us to see the effects on logical states rotated (spread) across the Hilbert space. Indeed, in a cluster of length $n$, the rotation applied to the logical qudit is given by $F^{n-1}$. The identity operation is therefore only applied to clusters whose lengths are multiples of $5$, as $F^4=\openone$. For qubits however, the identity operation is applied to all odd length clusters, as $F^2 \equiv H^2=\openone$, where $H$ is the Hadamard operation. For convenience, we consider measurement outcomes corresponding to the state $\ket{+}$ being obtained. A logical state $\ket{\psi}_d$ in a Hilbert space of $d$ dimensions can be parameterized by the Hurwitz parameterization \cite{Hurwitz}. Using angles $\theta_k \in [0,\pi/2]$ and $\phi_k \in [0,2 \pi)$ for $k=1,..,d-1$ we can write $\ket{\psi}_d=\sum_{i=0}^{d-1}c_i \ket{i}$,
where the coefficients $c_i$ are given by $c_0 = \cos \theta_1$, $c_j=(\Pi^{j}_{k=1}\sin\theta_k)\cos\theta_{j+1}e^{i\phi_j}~(0<j<d-1)$ and $c_{d-1}=(\Pi^{d-1}_{k=1}\sin\theta_k)e^{i\phi_{d-1}}$ \cite{hurwitzpara}.  
\begin{figure}[b]
\centerline{
\psfig{figure=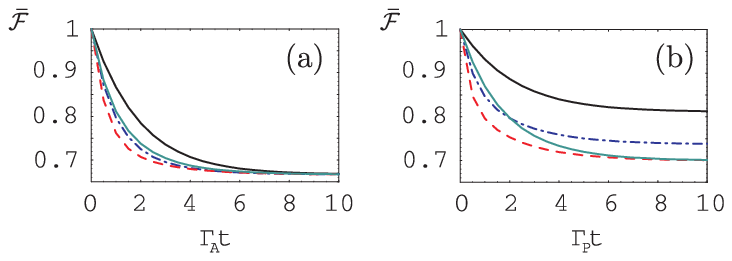,width=8cm}}
\caption{Fidelity decay for arbitrary single qudits with $d=2,~3,~4$ and $2 \otimes 2$ when affected by AD {\bf (a)} and PD {\bf (b)}. In both panels the upper solid, lower solid, dash-dotted and dashed lines correspond to $d=2,~2 \otimes 2,~3$ and $4$ respectively.}
\label{figu3}
\end{figure}
A pure state $|{\psi}\rangle_d$ representing the logical qudit will loose purity as it propagates across a linear cluster state under the influence of AD and PD. The rescaled interaction times $\Gamma_i t$ determine how fast purity is lost. In general ${\cal F}=\sqrt{\mbox{}_{d}\expect{\psi}{\varrho}{\psi}_{d}}={\cal F}(\{\theta_k\}_{k=1}^{d-1},\{\phi_k\}_{k=1}^{d-1}, \Gamma_i t)$, where $\varrho$ represents the mixed state of the logical qudit at the end of the cluster. In order to determine the behavior of the fidelity for an arbitrary state $\ket{\psi}$ one must average it over all angle sets $\{\theta_k\}$ and $\{\phi_k\}$ representing the configuration space $\Omega$ at each instant of time $\Gamma_i t$. This allows us to find the mean fidelity $\bar{\cal F}(\Gamma_i t)$ by using the multi-dimensional integral $\int_{\Omega}{\rm d}\nu$ with ${\rm d}\nu={(1/{\rm V}(\Omega))}\prod_{k=1}^{d-1}\cos\theta_k(\sin\theta_k)^{2k-1}{\rm d}\theta_k{\rm d}\phi_k$~\cite{Hurwitz} and
the total volume for the manifold of pure states given by ${\rm V}(\Omega)=[\pi^{d-1}/(d-1)!]$. This gives the mean fidelity $\bar{\cal F}(\Gamma_i t)=\int_{\Omega}~{\cal F}(\{\theta_k\}_{k=1}^{d-1},\{\phi_k\}_{k=1}^{d-1},\Gamma_i t){\rm d}\nu$, at each instant of time $\Gamma_i t$. We are now in a position to provide a quantitative picture of how propagated information is affected on average as the cluster is exposed to decoherence. In Fig. \ref{figu3} {\bf (a)} and {\bf (b)} we show respectively the effect of AD and PD on arbitrary single logical qudits encoded onto physical qudits for $d=2,~3$ and $4$. The average fidelity decays faster as the dimension increases. We also consider a $d=4$ qudit encoded onto an entangled pair of qubits $\ket{\psi}_{2\otimes{2}}$, where each is individually affected by AD and PD as shown in Fig. \ref{figu1a} {\bf (b)}. We use the definition $\ket{\psi}_{2\otimes{2}}=\sum_{i=0}^{3}c_i\ket{\tilde{i}}_{12}$, where $\tilde{i}$ is the binary expression for the integer $i$, the subscripts $1$ and $2$ label the qubits and $\{ c_i \}$ is that of a $d=4$ qudit. Evidently the entangled pair has a slower decay than that of the $d=4$ qudit in both decoherence models.
\begin{figure}[t]
\centerline{
\psfig{figure=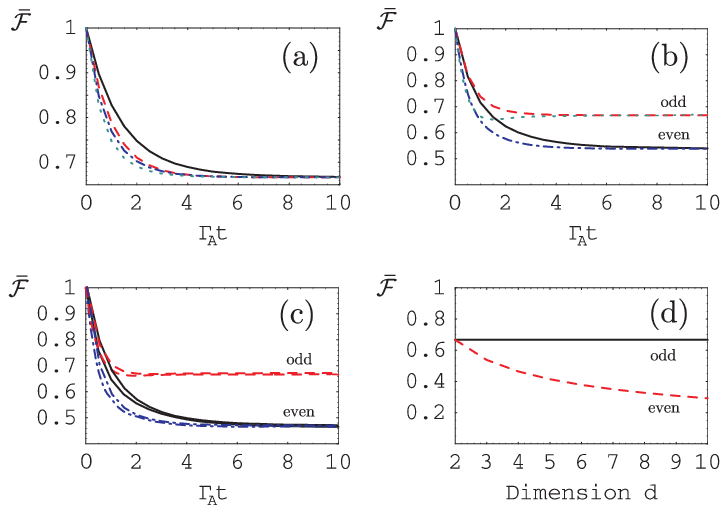,width=8cm}}
\caption{Average fidelities of AD-affected qudits propagated along qudit cluster states of lengths $n=2 \to 5$. In {\bf (a)}, {\bf (b)} and {\bf (c)} the solid, dashed, dash-dotted and dotted lines correspond to $n=2,~3,~4$ and $5$ length clusters respectively. {\bf (a)}: AD for $d=2$. {\bf (b)}: AD for $d=3$. {\bf (c)}: AD for $d=4$ compared with $2 \otimes 2$ (top lines at each $n$). {\bf (d)}: Even and odd length average fidelity for $\Gamma_{\rm A} t\to \infty$. The solid (dashed) line corresponds to odd (even) lengths.}
\label{figu4}
\end{figure}

To find out if the fidelity behaviors discussed above carry over to information transfer, we consider in Fig. \ref{figu4} the average fidelities of arbitrary encoded logical qudits propagated along qudit cluster states of lengths $n=2 \to 5$ when AD individually affects the physical qudits. Comparing Fig. \ref{figu4} {\bf (a)}, {\bf (b)} and {\bf (c)} corresponding to dimensions $d=2,~3$ and $4$ respectively, it becomes clear that there is a splitting effect seen only for dimensions $d>2$, where clusters of even length suffer a more pronounced fidelity decay than those of odd length. In addition to this, for even lengths there is a noticeable drop in the fidelities as $\Gamma_{\rm A} t \to \infty$ for increasing dimension. In Fig.~\ref{figu4} {\bf (d)} we show these final even and odd fidelity values against increasing dimension. In Fig.~\ref{figu4} {\bf (c)}, regardless of the splitting effects, we see that a $d=2\otimes{2}$ linear cluster always outperforms a $d=4$ one. 

The reason for the splitting in odd and even lengths is the following. The state of a logical qudit propagated across a cluster of length $n\ge 2$ in the limit $\Gamma_{\rm A} t \to \infty$ becomes equal to that of the final physical qudit in the same limit; \ie it becomes $\ket{0}\bra{0}$. When the fidelity is taken and averaged over the configuration space, we obtain
$\bar{\cal F}(\Gamma_{\rm A} t\to \infty)=\int_{\Omega}{|\bra{0}F^{n-1}\ket{\psi}_{d}|} {\rm d} \nu$. 
\begin{figure}[b]
\centerline{
\psfig{figure=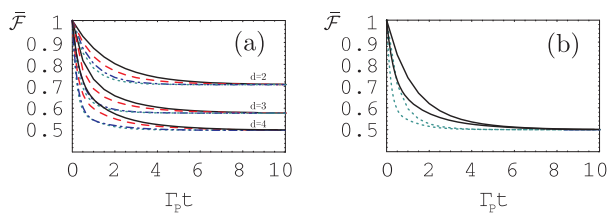,width=8cm}}
\caption{Average fidelities of PD-affected qudits propagated along qudit cluster states of lengths $n=2 \to 5$. The solid, dashed, dash-dotted and dotted lines correspond to $n=2,~3,~4$ and $5$ respectively. {\bf (a)}: The top, middle and bottom four lines correspond to dimensions $d=2,3,4$. {\bf (b)}: Comparison between $d=4$ and $d=2 \otimes 2$. The top lines always correspond to an entangled pair for each $n$.}
\label{figu5}
\end{figure}
For odd length clusters (even powers of $F$) we have $(F^{\dag})^{n-1}\ket{0}=\ket{0}$ because $F^2=\sum_{k=0}^{d-1}\ket{-k}\bra{k}$ and for even length clusters (odd powers of $F$) we have $(F^{\dag})^{n-1}\ket{0}=\ket{+}$. Therefore for odd lengths, only the $\ket{0}$ entry of the logical qudit state $\ket{\psi}$ takes part in the overlap. When the average is performed over the configuration space only $\theta_1$ is averaged. In the even length case we have the integral $\bar{\cal F}(\Gamma_{\rm A} t\to \infty)=(1/\sqrt{d})\int_{\Omega}|\sum_{i=0}^{d-1}c_i|{\rm d}\nu$. For qubits this gives $2/3$, which matches the odd length integral. 

For PD, no splitting effects arise because as $\Gamma_{\rm P} t \to \infty$ the final logical qudit state becomes $(1/d)\openone$, giving a fidelity of $\bar{\cal F}(\Gamma_{\rm P} t\to \infty)=1/\sqrt{d}$ for all lengths $n\ge2$. For $n=1$, the final state of the qudit (entangled pair) as $\Gamma_{\rm P} t \to \infty$ becomes $\varrho=\sum_{i=0}^{d-1}|c_i|^2\ket{i}\bra{i}$ ($\sum_{i=0}^{3}|c_i|^2\ket{\tilde{i}}\bra{\tilde{i}}$) leading to higher final fidelity values than in the case of arbitrary $n$, as can be seen by comparing Fig. \ref{figu3} {\bf (b)} with Fig. \ref{figu5} {\bf (a)}. In Fig. \ref{figu5} {\bf (a)} we show the average fidelity decay for arbitrary qudits propagated along linear clusters of length $n=2 \to 5$ for dimensions $d=2,3$ and $4$. Here it is evident that as the dimension increases, the fidelity decays becomes faster for all length clusters. For clarity, we separate the $d=2 \otimes 2$ case and compare it with the $d=4$ case in Fig. \ref{figu5} {\bf (b)}. For each $n$ the $d=2 \otimes 2$ cluster outperforms the $d=4$ qudit cluster. 

To explain the robustness of the entangled qubit pair one needs to consider how the environment acts on each physical cluster qudit. Due to the nature of the PD environment which scatters off each qudit system in the cluster, the terms $\varrho_{nm}(t)\ketbra{n}{m}~(n\neq m)$ of the density matrix decay faster for larger values of $(n-m)$. One can see this from the time dependence of these terms by using Eq. (\ref{P}) in the Kraus operator evolution to obtain the relation $\varrho_{nm}(t)\ketbra{n}{m}=\varrho_{nm}(0)e^{-\frac{1}{2}\Gamma_P t (n-m)^2}\ketbra{n}{m}$. The behaviors shown above suggest that it is best to restrict logical qudit simulations to lower levels in smaller dimensional physical qudits that are entangled, rather than using the same dimension for the physical qudits in the cluster. In this way we exclude faster decay terms due to larger differences in the levels between low and high number states. Thus, when PD is considered, $d=2\otimes2$ will always outperform $d=4$.

For AD the elements of $\varrho$ decay in favor of $\varrho_{00}(t)\ketbra{0}{0}$ ($\varrho_{00}(t)\to{1}$ as $\Gamma_A t\to\!\infty$). 
In these dynamics, the slowly decaying elements for $d=4$ are $\varrho_{00}(t)\ketbra{0}{0},\varrho_{01}(t)\ketbra{0}{1}$, $\varrho_{10}(t)\ketbra{1}{0}$. For $d=2\otimes{2}$, $\varrho_{00}(t)\ketbra{\tilde{0}}{\tilde{0}},\varrho_{01}(t)\ketbra{\tilde{0}}{\tilde{1}},\varrho_{10}(t)\ketbra{\tilde{1}}{\tilde{0}},\varrho_{02}(t)\ketbra{\tilde{0}}{\tilde{2}}$ and $\varrho_{20}(t)\ketbra{\tilde{2}}{\tilde{0}}$ are the slowly-decaying ones. The last two elements give an additional contribution to the fidelity with respect to the $d=4$ case. Their presence is understood by inspecting $\varrho$ in the qubit basis where 
we can see that the last two terms have the same total energy as the second and third terms and are identically affected by the AD environment, which cannot distinguish between them. 
For higher dimensions, similar considerations can be made for PD and AD. We conjecture that, based on the arguments described above, $d=2\otimes..\otimes{2}$ systems will have slower average fidelity decay than their $d$-dimensional equivalents.

\subsection{Encoded information transfer}
From the analysis of information flow it seems that moving to higher dimensions greatly decreases the transfer quality, characterized by the state fidelity, when decoherence is present. As the one-way model is based on the ability of transferring information across linear subclusters comprising the entangled resource, this sets a serious hindrance on the use of higher-dimensional systems. However, the previous analysis did not exhaust the possibilities offered by the employment of $d$-dimensional elements. Some advantages could come by encoding a logical qubit within the logical qudit being propagated. The average fidelity decays shown so far cover the entire Hilbert space for a particular $d$. Qubits encoded in these spaces do not necessarily make use of the full space and some advantage could be obtained by (in some sense) {\it hiding} the information from the environment. 

\begin{figure}[b]
\centerline{
\psfig{figure=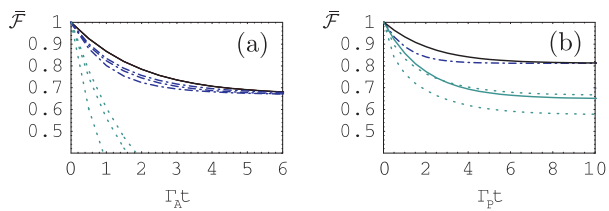,width=8cm}}
\caption{Average fidelities of qubits encoded into decoherence-affected single qudits. The upper solid lines correspond to $d=2$ and $G$ encodings for $d=3$ and $4$ in both {\bf (a)} and {\bf (b)}. In {\bf (a)} we consider AD. The dash-dotted lines correspond to an $L$-encoded qubit for $d=2 \otimes 2,~3$ and $4$ (from top to bottom). The dotted lines correspond to $M$ encoding for $d=2 \otimes 2$, $T$ encoding for $d=3$ and $4$ from top to bottom respectively. In {\bf (b)} we consider PD. The $G$ and $T$ encodings for $d=3$ and $4$ match the $d=2$ case. The dash-dotted line is for an $O/M$-encoded qubit in $d=2\otimes 2$. The dotted lines correspond to $E$ encoding for $d=3$ and $4$ (from top to bottom). The lower solid line is for $L$ encoding in $d=2\otimes 2$.}
\label{figu6a}
\end{figure}

\begin{figure}[t]
\centerline{
\psfig{figure=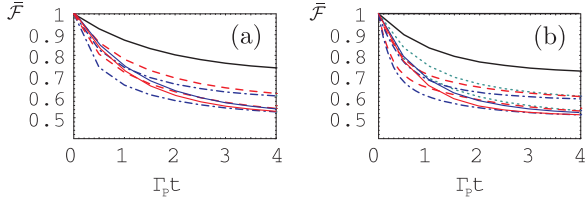,width=7.7cm}}
\caption{Average fidelities of PD-affected encoded qubits propagated along qudit cluster states of lengths $n=2$ and $3$. The top solid lines correspond to $d=2$, the dashed lines to a $G$ encoding for $d=3$ and $4$ (from top to bottom), the dash-dotted lines to $L$ encoding for $d=3$ and $4$ (from top to bottom), and the dotted lines to $T$ encoding for $d=3$ and $4$ (from top to bottom). {\bf (a)}: $n=2$ length cluster. $T$ encodings match up with $G$ encodings for $n=2$. For $d=2\otimes 2$, the middle (bottom) solid line corresponds to $L$ ($O/M$) encoding. {\bf (b)}: $n=3$ length cluster. For $d=2 \otimes 2$, the middle (bottom) solid line corresponds to $L/M$ ($E$) encoding.}
\label{figu7}
\end{figure}

In order to introduce the encoding techniques, we look back at the Hurwitz parameterization. A qubit state $\ket{\psi}_2 \in {\cal H}_2$ can be described simply as a state within the subspace of a $d=3$ dimensional Hilbert space ${\cal H}_3$ where $\theta_2=0$. In general, any $d'$-dimensional state $\ket{\psi}_{d'}$ ($d'<d$) can be described as a state within a particular subspace of a $d$-dimensional Hilbert space ${\cal H}_d$. We can thus take a state $\ket{\psi}_{d'} \in {\cal H}_d$ and use a unitary transformation $\Lambda_d$ to encode the state into the entire Hilbert space. To apply an operation $\chi_{d'}\in{\cal H}_{d'}$ to a state encoded in a larger Hilbert space, we use the transformation
\begin{equation}
\label{chi1}\chi_{d}=\Lambda_d~ \tilde{\chi}_{d'} \Lambda_d^{\dag},~~\chi_d \in {\cal H}_d
\end{equation}
with $\tilde{\chi}_{d'}={\chi}_{d'}\oplus{R}$ and $R$ is a $(d-d')\times(d-d')$ matrix with arbitrary phase factors along its diagonal. These phases can be used to simplify the encoded operation $\chi_d$. We want to encode qubits ($d'=2$) into higher dimensions ($d \ge 3$) and manipulate them using $d$-dimensional one-way QC. We consider encoded states $\ket{\psi_{\cal E}}_d$ given by
\begin{equation}
\ket{\psi_{\cal E}}_d=\Lambda_d(a\ket{0}+ b\ket{1}),~~\ket{\psi_{\cal E}}_d,\Lambda_d \in {\mathcal H}_d,
\end{equation}
where $a=\cos\theta_1$ and $b=\sin \theta_1 e^{i \phi_1}$ \cite{encodeeg}. 
Even though we have considered many other types of encodings, for clarity we show the performances of only the best and worst encodings found for each dimension, under AD and PD. In Fig.~\ref{figu6a} we show the effect of encoding logical qubits into single physical qudits. In our notation, ${\cal E}$ denotes the encoding type and $d$ gives the dimension for which the encoding is used. The encodings are given in Table \ref{tab1}.
\begin{table}[t]
\begin{ruledtabular}
\begin{tabular}{|c|l|}\hline
Encoding &\hskip3cm{St}ate\Large{\phantom{A}} \\ \hline\hline
G-Ground & $\ket{\psi_G}_3=\ket{\psi_G}_4=a\ket{0}+ b \ket{1}$. \Large{\phantom{A}}\\ \hline
T-Top & $\ket{\psi_T}_3=a\ket{1}+ b\ket{2},~
\ket{\psi_T}_4=a\ket{2}+ b\ket{3}$.~~~\Large{\phantom{A}} \\ \hline
L-Lopsided & $\ket{\psi_L}_d=a\ket{0}+\frac{b}{\sqrt{d-1}}\sum^{d-1}_{j=1}\ket{j}$~(d=3,4), \Large{\phantom{A}}\\
& $\ket{\psi_L}_{2\otimes{2}}=a\ket{\tilde{0}}+ \frac{b}{\sqrt{3}}(\ket{\tilde{1}}+\ket{\tilde{2}}+\ket{\tilde{3}})$. \\ 
\hline
O-Outside & $\ket{\psi_O}_{2\otimes{2}}=a\ket{\tilde{0}}+ b\ket{\tilde{3}}.$ \Large{\phantom{A}} \\ \hline 
M-Middle & $\ket{\psi_M}_{2\otimes{2}}=a\ket{\tilde{1}}+ b\ket{\tilde{2}}.$ \Large{\phantom{A}} \\ \hline
E-Equal & $\ket{\psi_E}_d=\frac{1}{\sqrt{d}}\sum^{d-1}_{n=0}(a+\omega^nb)\ket{n}$~~(d=3,4), \Large{\phantom{A}}\\
& $\ket{\psi_E}_{2\otimes{2}}=\frac{1}{2}\sum^3_{{n}=0}(a+\omega^{n}b)\ket{\tilde{n}}.$  \Large{\phantom{A}} \\ \hline
\end{tabular}
\end{ruledtabular}
\caption{Encodings used in the analysis.}
\label{tab1}
\end{table}
One can see in Fig.~\ref{figu6a} {\bf (a)} that for AD, $G$ encoding is the best for encoding qubits into single qudits as the fidelity-decays match exactly that of a single qubit. The worst encoding for AD is given by $T$. This is because in the limit $\Gamma_{\rm A} t\to \infty$ the final state of the qudit becomes $\ket{0}\bra{0}$ for all dimensions and therefore the average fidelity $\bar{\cal F}(\Gamma_{\rm A} t\to \infty)=0$. The next best encoding is given by $L$, for any dimension. In Fig.~\ref{figu6a} {\bf (b)} we show the PD case, where one can see that $G$ and $T$ represent the best encoding for qubits into single qudits. The next best encodings are $L$, for both $d=3$ and $4$, and $O/M$ for $d=2 \otimes 2$, the latter performing significantly better than that for $d=3$ and $4$. Moreover, it is known that to transmit qubits through qudit channels in presence of the AD and PD models we have considered, the best encoding is given by using the two lowest and two contiguous states respectively~\cite{Amosov}. The results shown here agree with this finding. We now investigate to see whether this feature holds true also for information-propagation along cluster states.
\begin{figure}[t]
\centerline{
\psfig{figure=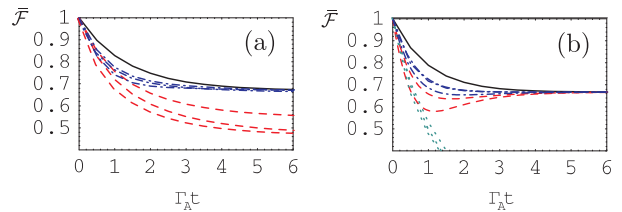,width=8cm}}
\caption{Average fidelities of AD-affected encoded qubits propagated along qudit cluster states of lengths $n=2$ and $3$. Solid lines are for $d=2$ in all panels. {\bf (a)}: $n=2$ length cluster. dash-dotted lines represent $L$ encoding for $d=2\otimes 2,~3$ and $4$ (from top to bottom). Dashed lines represent $G$ encoding for $d=3$ and $4$, and $M/O$ encoding for $d=2 \otimes 2$ (from top to bottom). {\bf (b)}: $n=3$ cluster. dash-dotted lines represent $L$ encoding for $d=2\otimes 2,3$ and $4$ (from top to bottom). Dashed lines represent $G$ encoding for $d=3$ and $4$ (from top to bottom). Dotted lines represent $T$ encoding for $d=3$ and $4$, and $M$ encoding for $d=2 \otimes 2$ (from top to bottom).}
\label{figu6}
\end{figure}
In Fig.~\ref{figu6} we show average fidelities of encoded qubits propagated across AD-affected qudit clusters of lengths $n=2$ and $3$. We have also checked the case of $n=4$ ($n=5$), which has similar behavior to $n=2$ ($n=3$). No encoding surpasses the qubit cluster state propagation, regardless of $n$. The next best encodings come from $d=2 \otimes 2$, where $L$ encoding is the best. The worst encodings are $G$ ($M/O$) for even-length clusters and $T$ ($M$) for odd-length ones with $d=3$ and $4$ ($2\otimes 2$). In Fig.~\ref{figu7} we show average fidelities of encoded qubits propagated across PD-affected qudit clusters of lengths $n=2$ and $3$. 
Evidently, no encoding surpasses the propagation through qubit clusters. The next best encoding for $d=3,4$ is $T$ ($\forall{n}>2$) 
while $L$ is always the worst.
\subsection{Encoded gate simulation}
\begin{figure}[h]
\centerline{
\psfig{figure=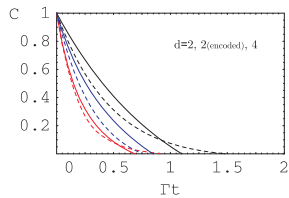,width=3.72cm}}
\caption{Entanglement in decohered BBB$_3$-produced cluster states. In all graphs the solid (dashed) lines correspond to PD (AD). We show $c(\varrho)$ for $d=2$ (top two curves), $c_{qp}(\varrho)$ for $d=4$ (bottom two curves) and $c(\varrho)$ for 2 qubits encoded in the lowest levels of $d=4$ logical qudits and propagated through BBB$_3$ (middle two curves).}
\label{figu9}
\end{figure}

We now study gate operations on qubits encoded within logical qudits in a $d$-dimensional cluster state. Consider the entangling gate BBB$_2$ (see Fig.~\ref{tfig1} (c)) with two qubits labeled $\ket{q_{1,2}}$ encoded in two logical qudits, labeled $\ket{Q_{1,2}}$ respectively. Assume the two qubits are decoded just before an entangling gate is simulated on the logical qudits.
Let $\ket{q_1}=a\ket{0}+b\ket{1}$ and $\ket{q_2}=c\ket{0}+d\ket{1}$. 
Within the unencoded subspace of ${\mathcal H}_3$ we have the entangling gate $E_{12}=|0\rangle_1\langle 0| \otimes {\openone}_{2} + |1 \rangle_1 \langle 1| \otimes (|0 \rangle_2 \langle 0|+ e^{i 2\pi/3}|1 \rangle_2 \langle 1|)$. Two applications of $E_{12}$ (and local rotations) are required to implement $S_{12}$ for $d=2$ (see Eq.~(\ref{Sabdef}))~\cite{Hammerer,NC}. However, together with Hadamard and $R_z^{\gamma}$ rotations, it is sufficient for universality. 
To understand how the entangling capabilities of BBB$_2$ and BBB$_3$ are affected by decoherence, we can take the case when both logical input qudits are in $\ket{+}$. Under ideal conditions, both BBB$_2$ and BBB$_3$ will create bipartite states LU equivalent to maximally entangled states. We have already investigated entanglement decay for BBB$_2$ in higher dimensions in Section III. Indeed, the entanglement generated between two logical qudits in the state $\ket{+}$ through BBB$_2$ is the same as that for a $2$-qudit cluster state. We are therefore interested in the amount of entanglement generated by BBB$_3$ under decoherence. We would also like to study how the decoherence effects on the entanglement generated by BBB$_3$ between two qubits in $\ket{+}$ encoded in the two lowest states of logical qudits of $d=4$. In the ideal case described above, a $n=2$ qubit state LU equivalent to a maximally entangled state is generated. In Fig.~\ref{figu9} {\bf (a)} we show the concurrence decay for BBB$_3$ when $d=2$ and normalized quasi-concurrence when $d=4$. We also show the case when two qubits are encoded into the two lowest states and sent through BBB$_3$ for $d=4$. As the dimension increases, the proportion of the maximum achievable entanglement decays faster for both AD and PD. When two qubits are encoded into two qudits ($d=4$), one can see that entanglement generated by the BBB$_3$ gate, in terms of concurrence, decays much faster than in the qubit case. 
\section{Conclusions}
We have investigated the extension of the one-way model for QC to $d$-dimensional systems by providing a thorough analysis of entanglement properties, information transfer and gate simulation under environmental influence. Such an extension, performed so far without reasonable justification, appears not to provide any advantage with respect to the {\it standard} qubit-based one-way model, when global properties of the entanglement resource are used in order to quantify the performances of a given protocol. Indeed, our study also reveals the previously overlooked superiority of a resource built out of pairs of entangled two-level systems with respect to higher-dimensional elementary systems. Furthermore, this work suggests (for the models considered) the exclusion of the use of $d$-dimensional systems in measurement-based QIP as a tool for protecting information from the effects of environmental noise. 
\acknowledgments 
We thank V. Vedral for discussions, DEL, the Leverhulme Trust (ECF/40157), UK EPSRC and KRF (2003-070-C00024) for financial support.
\appendix
\section*{Appendix}
\begin{figure}[b]
\centerline{
\psfig{figure=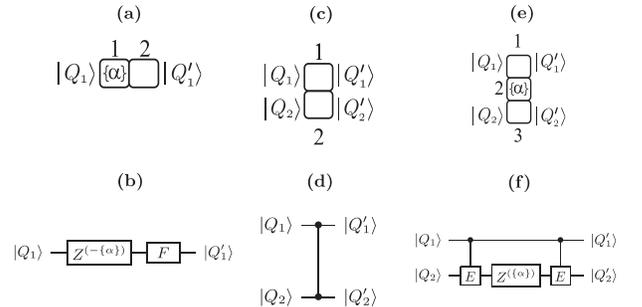,width=8cm}}
\caption{{\bf (a)}, {\bf (c)} $\&$ {\bf (e)} show the layouts of BBB$_1$, BBB$_2$ and BBB$_3$. {\bf (b)}: The operation simulated on logical qudit $\ket{Q_1}$, when physical qudit $1$ is measured in the $B_1(\{\alpha\})$ basis and $s_1=0$ is obtained. {\bf (d)}: The $S_{12}$ gate simulated by BBB$_2$ on two logical qudits $\ket{Q_1}$ and $\ket{Q_2}$. {\bf (f)}: The quantum circuit corresponding to the operation BBB$_3$ with $\{\alpha\}$ satisfying the conditions in the text, when $s_2=0$. Here, $C-E=F_2S_{12}F_2$.}
\label{tfig1}
\end{figure}
Our diagrammatic notation is such that each physical qudit is represented by a rounded-square, with empty ones denoting unmeasured input/output qudits. The angle set $\{ \alpha \}$ inside the $i$th qudit symbol identifies the basis $B_i(\{\alpha\})=\{Z^{(\{ \alpha \})} \ket{+_j} \}_{j=0}^{d-1}$ in which that qudit is measured. Here, $Z^{(\{\alpha \})}=\sum_{k=0}^{d-1} e^{i \alpha_k} \ketbra{k}{k}$ is completely defined by the set of angles $\{ \alpha\}= \{ \alpha_k \in [0,2\pi) \}$ and $s_i \in \{ 0,..,d\}$ is the corresponding measurement outcome. 
The smallest cluster state consists of two qudits and can be used to simulate a unitary operation on one logical qudit encoded on a physical cluster qudit~\cite{Zhou1}, as shown in Fig.~\ref{tfig1} {\bf (a)}. We denote this configuration as BBB$_1$. The operation simulated by BBB$_1$ when a measurement on qudit $1$ gives outcome $s_1=0$ is shown in Fig.~\ref{tfig1} {\bf (b)}. There, $Z^{(-\{\alpha \})}$ represents the rotation carried out on the logical qudit and $F$ is the quantum Fourier transform in $d$-dimensions, given by $F={d}^{-1/2}\sum_{j,k=0}^{d-1}\omega^{jk}|j\rangle\langle k|$. Due to the probabilistic nature of the simulation, it is necessary to apply a decoding operator ${\cal D}(s_1)=X^{s_1}$ to qudit $2$, found via the relations $XZ=\omega ZX$, $FZ=XF$ and $FX=Z^{\dag}F$ \cite{Zhou1}. Using the same layout, with two encoded qudits $\ket{Q_{1,2}},$ 
the operation $S_{12}$ in Eq.~(\ref{Sabdef}) is simulated as shown in Fig.~\ref{tfig1} {\bf (d)}. We denote this configuration as BBB$_2$. Finally, in Fig.~\ref{tfig1} {\bf (e)} we have qudits $1$ and $3$ embodying the input logical qudits and a measurement is performed on $2$ in the $B_{2}(\{\alpha \})$ basis. This pattern simulates the operation
$T_{13}(s_2)={d}^{-1/2}\sum_{k,l,j=0}^{d-1}\omega^{j(k+l-s_2)}e^{-i \alpha_j} |k,l \rangle_{1,3} \langle k,l|$.
This is unitary only when the set $\{\alpha\}$ satisfies $|\sum_{j=0}^{d-1}\omega^{j(k+l)}e^{-i \alpha_j}|^{2}=d$. The index $j$ of $\alpha_j$ follows a modulo-$d$ arithmetic. Additionally when ${d}^{-1} \sum_{j=0}^{d-1}\omega^{j(k+l)}e^{-i \alpha_j}=e^{i \alpha_{-(k+l)}},\forall{l},k\in\{0,..,d-1 \}$, we have for $s_2=0$ the unitary transformation $U_{13}={d}^{-1/2}\sum_{k,l=0}^{d-1}e^{i \alpha_{-(k+l)}} |k,l\rangle_{1,3} \langle k,l|$.
This corresponds to the operation in Fig.~\ref{tfig1} {\bf (f)}. 
Using this set of BBB's, we can construct more complicated configurations using a simple concatenation technique \cite{RBH,Tame1} which holds true for any $d$-dimensional qudit cluster state. To find the form of the decoding operators ${\cal D}$ to apply to the output logical qubit of a particular concatenated cluster configuration, we can use the relations~\cite{Zhou1} $Z^{(\{ \alpha\})}Z=Z Z^{(\{ \alpha\})},~Z^{(\{ \alpha\})}X=X Z^{(\{ \alpha'\})}~[\alpha'_j=\alpha_{j-1}],~FZ=XF,~FX=ZF,~S_{12}(X^{x_1}Z^{z_1})_1(X^{x_2}Z^{z_2})_2 = (X^{x_1}Z^{z_1-x_2})_1(X^{x_2}Z^{z_2-x_1})_2 S_{12}$.
The second relation implies the use of adaptive measurements in the simulations, similarly to the qubit case~\cite{RBH,Tame1}. For prime dimensions, universal quantum computation can be achieved with the set of $d+1$ single qudit rotations $\{ Z^{(\{ \alpha\})}, X^{(\{ \alpha\})}, Z(X^{(\{ \alpha\})})^{k}\},~k=1,..,d-1$ and the two-qudit gate $S_{12}$, where $X^{(\{ \alpha\})}=FZ^{(\{ \alpha\})}F^{\dag}$. Finding the corresponding universal sets in the case of any dimension is more involved~\cite{Zhou1}.


\begin{thebibliography}{99}
\bibitem{encodeeg} One could encode the qubit state across a qutrit, using $\Lambda_3=1 \oplus H$
to obtain $\ket{\psi_{\cal E}}_3= a\ket{0}+ b(\ket{1}+\ket{2})/\sqrt{2}$. A rotation on this encoded qubit given by $\chi_2=R^{\gamma}_z\equiv \ketbra{0}{0}+ e^{i \gamma} \ketbra{1}{1}$ uses 
$\tilde{\chi}_2=\ketbra{0}{0}+ e^{i \gamma} \ketbra{1}{1}+ e^{i \delta} \ketbra{2}{2}$ setting $\delta=\gamma$ to give $\chi_3=\ketbra{0}{0}+ e^{i \gamma} \ketbra{1}{1}+ e^{i \gamma} \ketbra{2}{2}$. For $\chi_2=H$ we get $\chi_3=\cos^2(\pi/8)\openone +\cos^2(3 \pi/8) \ketbra{0}{0}+[\ketbra{0}{1}+\ketbra{0}{2}-2 \cos^2(3 \pi/8)\ketbra{1}{2} + h.c.]/2$, by setting $\delta=0$. This can be applied to any $d \ge 3$ and $\Lambda_d$ to find equivalent operations for $R_z^{\gamma}$ and $H$. These are sufficient to carry out any qubit rotation \cite{RBH,NC}. We can simulate any rotation on a $d$-dimensional logical qudit using $d$-dimensional cluster states \cite{Zhou1}. Thus, arbitrary rotations can be simulated on qubits encoded within logical qudits.
\end{thebibliography}
\end{document}